\documentclass[prl, superscriptaddress, showpacs, floatfix,
twocolumn]{revtex4}
\usepackage{amssymb}
\usepackage{graphicx}
\usepackage{amsmath}
\usepackage{amsfonts}

\newcommand{\ket}[1]{\ensuremath{|#1 \rangle}}

\newcommand{\kb}[1]{\ensuremath{| #1 \rangle \! \langle #1 |}}

\newcommand{\cb}[1]{\ensuremath{\| #1 \|_{\diamond}}}

\newcommand{\id}{\ensuremath{{\rm id} \, }}

\def\one{{\mathchoice{\rm 1\mskip-4mu l}{\rm 1\mskip-4mu l}{\rm 1\mskip-4.5mu l}{\rm
1\mskip-5mu l}}}
\def\ket#1{| #1 \rangle}

\def\kb#1#2{|#1\rangle\!\langle #2 |}

\def\C{{\cal C}}

\def\E{{\cal E}}
\def\F{{\cal F}}
\def\G{{\cal G}}

\def\M{{\cal M}}

\def\R{{\cal R}}

\def\Tr{{\mathrm{tr}}}

\newtheorem{theorem}{Theorem}

\newtheorem{definition}[theorem]{Definition}

\newtheorem{lemma}[theorem]{Lemma}


\begin{document}

\date{\today}
\title{Complementarity of Private and Correctable Subsystems \\ in Quantum Cryptography and Error Correction}
\author{Dennis Kretschmann}
\affiliation{Institut f\"ur Mathematische Physik, Technische
Universit\"at Braunschweig, Mendelssohnstra{\ss }e~3, 38106
Braunschweig, Germany} \affiliation{Quantum Information Theory
Group, Dipartimento di Fisica A.~Volta, Universit\`{a} di Pavia,
via Bassi 6, 27100 Pavia, Italy}
\author{David~W.~Kribs}
\affiliation{Department of Mathematics and Statistics, University
of Guelph, Guelph, ON, Canada, N1G 2W1} \affiliation{Institute for
Quantum Computing, University of Waterloo, ON Canada, N2L 3G1}
\author{Robert~W.~Spekkens}
\affiliation{Department of Applied Mathematics and Theoretical
Physics, Cambridge University, Cambridge, UK, CB3 0WA}

\begin{abstract}
We make an explicit connection between fundamental notions in
quantum cryptography and quantum error correction.
Error-correcting subsystems (and subspaces) for quantum channels
are the key vehicles for contending with noise in physical
implementations of quantum information-processing. Private
subsystems (and subspaces) for quantum channels play a central
role in cryptographic schemes such as quantum secret sharing and
private quantum communication. We show that a subsystem is private
for a channel precisely when it is correctable for a complementary
channel. This result is shown to hold even for approximate notions
of private and correctable defined in terms of the diamond norm
for superoperators.
\end{abstract}

\maketitle


In operator quantum error correction, a \emph{correctable
subsystem} for a
noise map is one upon which the action of the noise can be corrected \cite%
{KLP05,KLPL05}. \ Equivalently, it is one that merely suffers a
unitary change of representation and therefore does not decohere
at all \cite{KS06,Kni06}. \ A \emph{private subsystem} is the
extreme opposite: it is one that completely decoheres under the
action of the noise in the sense that no information about the
state of the subsystem remains at the output of the map
\cite{BRS04}. \ This concept is very useful in quantum
cryptography. \ For instance, it finds application in the context
of private quantum communication schemes. \ If Alice encodes
quantum information using a secret key that she shares with Bob,
then Eve's ignorance of this key can be modelled as a noisy
channel. \ As an example, suppose Alice and Bob share a secret
classical key in the form of a random variable $X$ with
distribution $p$ which they use to select a unitary from a set
$\{U_{x}\}$ to implement on a
system prior to transmitting it. Then Eve's description of the system is $%
\mathcal{E}(\rho )=\sum_{x}p(x)U_{x}\rho U_{x}^{\dag }.$ The
private subsystems of this channel are precisely the subsystems
about which Eve obtains no information \cite{AMTW00,BRS04,BHS05}.
\ Another cryptographic application is quantum secret sharing
\cite{CGL99,CGS02}. \ Suppose a system is mapped by a channel
$\mathcal{C}$ to $n$ systems, distributed among $n$ parties. \ How
can one encode quantum information into this system in such a way
that any set of parties with less than $k$ members can learn
nothing about it? \ The answer is that it must be encoded into
subsystems that are private for the reduction of $\mathcal{C}$ to
any $k-1$ or fewer parties.

Finding the private subsystems for an arbitrary map is therefore a
problem with significant applications in quantum cryptography. \
It is in fact the counterpart of one of the central problems in
quantum error correction -- finding the correctable subsystems for
an arbitrary noise map. \ This problem, which encompasses that of
finding the error-correcting subspace codes
\cite{Sho95a,Ste96a,Got96,BDSW96a,KL97a} and the decoherence-free
subspaces and noiseless subsystems \cite%
{PSE96,DG97c,ZR97c,LCW98a,KLV00a,Zan01b,KBLW01a}, has been the
subject of intensive investigations of late
\cite{HKL04,Pou05,SL05,Bac05,CK06,Kni06,KS06,KlaSar06,BKK07a,BNPV07}.
On the other hand, almost no work has been done on the private
subsystem problem. \ At first glance then, one might expect the
road to progress to be a long one. \ However, the duality of the
two problems provides a shortcut. \ Indeed, a central result of
our paper is that the private subsystems for a map are simply the
correctable subsystems for a complementary map, where the notion
of complementarity of maps is the one introduced in \cite{DS03}. \
Because it is straightforward to obtain the complements of a map,
it follows that all the techniques and progress on finding
correctable subsystems can be immediately appropriated for the
problem of finding private subsystems.

The implication also holds in the opposite direction: the
correctable subsystems for a map are the private subsystems for a
complementary map.  Consequently, results from the field of
cryptography may also provide novel insights for error correction.

This duality between private and correctable in the case of
\emph{subspaces}
has already been used implicitly in previous work, such as \cite%
{CGL99,SP00}, where results from error correction are exploited to
derive conclusions for cryptography. Our result is therefore
likely to be intuitive to most quantum information theorists. \
Nonetheless, it is a surprisingly powerful tool. \ Indeed, many
well-known results in quantum information (and generalizations
thereof) can be derived as simple consequences of it.

In real-world applications, demanding perfect recovery or complete
decoherence of quantum information is too restrictive. We
therefore also consider approximate notions of error correction
and privacy, defined in terms of the diamond norm for
superoperators \cite{Kit97,AKN97}, which can be computed
algorithmically \cite{JKP07}. We demonstrate that if a subsystem
is approximately correctable (respectively private) for a map then
it is approximately private (respectively correctable) for a
complementary map. While an approximate version of the ideal
result is not unexpected, it is still surprising that simple
dimension-independent bounds can be derived. To accomplish this,
we make use of recently developed techniques from \cite{KSW06}.

We now describe preliminary notation and nomenclature. Given a
quantum system $S$ represented on a (finite-dimensional) Hilbert
space, also denoted by $S$, we say a quantum system $B$ is a
\textit{subsystem} of $S$ if there is a representation of $B$ such
that
$S=(A\otimes B)\oplus (A\otimes B)^{\perp }, $
where $A$ is also a subsystem of $S$. The {\it subspaces} of $S$
can be identified with subsystems $B$ for which $A$ is
one-dimensional. We adopt the convention that $A,B$ are subsystems
of $S$, and $A^{ \prime },B^{ \prime }$ are subsystems of $S^{
\prime }.$ We also adopt the convention that $\rho $ denotes a
density operator, and $\sigma $, $\tau $ denote arbitrary
operators. A subscript such as $\sigma_B$ refers to the subsystem
on
which the operator is defined. The set of linear operators on $S$ is denoted by $%
\mathcal{L}(S)$.

Linear maps on $\mathcal{L}(S)$, or ``superoperators'', can be
regarded as operators acting on the space $\mathcal{L}(S)$ with
the Hilbert-Schmidt inner product $(\sigma ,\tau
)={\mathrm{tr}}(\sigma ^{\dagger }\tau )$. We use the term
\emph{channel} to
mean a trace-preserving completely positive linear map $\mathcal{E}:\mathcal{%
L}(S)\rightarrow \mathcal{L}(S^{\prime })$ between Hilbert spaces $S$ and $%
S^{\prime }$. Such maps describe (discrete) time evolution of open
quantum systems in the Schr\"{o}dinger picture. The composition of
two
maps will be denoted by $\mathcal{E}\circ \mathcal{F}(\sigma )=\mathcal{E}(%
\mathcal{F}(\sigma ))$. A unitary channel $\mathcal{U}$ satisfies $\mathcal{U%
}^{\dag }\circ \mathcal{U}=\mathcal{U}\circ \mathcal{U}^{\dagger }=\mathrm{id%
}$, where $\mathrm{id}$ is the identity map, and is implemented by
a unitary operator $U$ via $\mathcal{U}(\sigma )=U\sigma
U^{\dagger }$. An isometric
channel $\mathcal{V}$ satisfies only $\mathcal{V}^{\dagger }\circ \mathcal{V}%
=\mathrm{id}$ and is implemented by an isometry $V$ via
$\mathcal{V}(\sigma
)=V\sigma V^{\dagger }$. Let $\mathcal{P}_{AB}$ be the map defined by $%
\mathcal{P}_{AB}(\sigma )=P_{AB}\sigma P_{AB}$ where $P_{AB}$ is
the projector onto the subspace $A\otimes B,$ and let
$\mathrm{id}_{B}$ be the
identity map on $\mathcal{L}(B)$. If we are given maps $\mathcal{E}_{A}$, $%
\mathcal{E}_{B}$ on the subsystems, as a notational convenience we write $%
\mathcal{E}_{A}\otimes \mathcal{E}_{B}$ both for the map on $\mathcal{L}%
(A\otimes B)$ and for the natural extension of the map to
$\mathcal{L}(S)$. \ Finally, the input and output spaces of
operators and superoperators will often be denoted by whether they
appear on the right or left of a conditional in the subscript,
e.g., $V_{BC|A}\mathpunct:A\rightarrow
B\otimes C$ and $\mathcal{E}_{C|AB}\mathpunct:\mathcal{L}(A)\otimes \mathcal{%
L}(B)\rightarrow \mathcal{L}(C).$ The absence of a conditional
implies
equality of input and output spaces, e.g. $\mathcal{E}_{A}\mathpunct:%
\mathcal{L}(A)\rightarrow \mathcal{L}(A).$

The norm distance $\cb{\cdot}$ that we use to quantify the
approximate cases of the main result is the {\em diamond norm} for
superoperators, originally introduced in the context of quantum
computing and error
correction \cite{Kit97,AKN97}. It is defined by $%
\left\Vert \mathcal{E}-\mathcal{F}\right\Vert _{\Diamond }:=
\sup_{k\geq 1}\left\Vert \mathrm{id}_{k}\otimes
(\mathcal{E}-\mathcal{F})\right\Vert _{1} $ where
$\mathrm{id}_{k}$ denotes the identity operation on the
complex-valued $(k \times k)$ matrices, and $\left\Vert \cdot
\right\Vert _{1}$ denotes the superoperator $1$-norm $\left\Vert
\mathcal{E}\right\Vert _{1}:= \sup_{\left\Vert \sigma \right\Vert
_{1}\leq 1}\left\Vert \mathcal{E}(\sigma
)\right\Vert _{1}$ where $\left\Vert \sigma \right\Vert _{1}=\mathrm{tr}%
|\sigma |.$ The diamond norm stabilizes in the sense that this
supremum is attained for $k$ equal to the dimension of the output
Hilbert space for the superoperator. (In fact, it is the dual of
the \emph{completely bounded norm}, $\left\Vert
\mathcal{E}\right\Vert _{\Diamond }=\left\Vert \mathcal{E}^{\dag
}\right\Vert _{cb}$ \cite{Pa}.)\ Channels $\mathcal{E}$ and
$\mathcal{F}$
are said to be $\epsilon $\emph{-close} if $\left\Vert \mathcal{E}-\mathcal{F%
}\right\Vert _{\Diamond }\leq \epsilon .$ \ If two channels are $\epsilon $%
-close, then the maximum probability of distinguishing the output
states of the channels, in an optimization over all input states
entangled with an ancilla of arbitrary dimension, is $1/2+\epsilon
/4.$ \ This follows from
the fact that $\frac{1}{2}+\frac{1}{4}\left\Vert \rho _{\mathcal{E}}-\rho _{%
\mathcal{F}}\right\Vert _{1}$ is the maximum probability of discriminating $%
\rho _{\mathcal{E}}=\mathrm{id}_{k}\otimes \mathcal{E}(\sigma )$ and $\rho _{%
\mathcal{F}}=\mathrm{id}_{k}\otimes \mathcal{F}(\sigma ),$ and
that the supremum over $\sigma$ in $\left\Vert
\mathcal{E}-\mathcal{F}\right\Vert _{\Diamond }=\sup_{k\geq
1}\sup_{\left\Vert \sigma \right\Vert _{1}\leq 1}\left\Vert
\mathrm{id}_{k}\otimes \mathcal{E}(\sigma )-\mathrm{id}_{k}\otimes \mathcal{F%
}(\sigma )\right\Vert _{1}$ captures the optimization.

We introduce the term \emph{deletion channel} for a channel that
has a 1-dimensional output space, that is, for all $\sigma _{B}\in
\mathcal{L}(B),$ $\mathcal{D}_{B^{\prime }|B}(\sigma
_{B})=\mathrm{tr}_{B}(\sigma _{B}) \,\omega _{B^{\prime }}$ for
some fixed $\omega _{B^{\prime }}.$ Note that the completely
depolarizing channel is a special case of a deletion channel where
$\omega _{B^{\prime }}\propto I_{B^{\prime }}.$\emph{\ \ }A
\emph{pure deletion channel} is one for which $\omega _{B^{\prime
}}$ is a pure state. \ The trace is a special case of a pure
deletion channel.

We now define what we mean by {\em private} and {\em correctable}
subsystems.
\begin{definition}
    \label{define:private}
        Given $\epsilon \geq 0$, we say $B$ is an $\epsilon $\emph{-private subsystem} for $%
        \mathcal{E}_{S^{\prime }|S}$ if there is a channel $\mathcal{M}_{A^{\prime
        }|A}$ and a deletion channel $\mathcal{D}_{B^{\prime }|B}$ such that%
        \begin{equation}
            \label{eq:private}
                \left\Vert \mathcal{E}_{S^{\prime }|S}\circ \mathcal{P}_{AB}-\mathcal{M}
                _{A^{\prime }|A}\otimes \mathcal{D}_{B^{\prime }|B}\right\Vert _{\Diamond
                }\leq \epsilon \, .
        \end{equation}
        If Eq.~(\ref{eq:private}) holds with $\epsilon = 0$, we call $B$ a {\em private subsystem}.
\end{definition}
This can be seen as an improved definition of $\epsilon $-private
relative to the one presented in \cite{BHS05}, because it
guarantees privacy even if the eavesdropper holds a purification
of the state. The term \textquotedblleft
completely\textquotedblright\ private was used in \cite{BRS04} to
describe private subsystems, but we drop this term here for
succinctness. In the $\epsilon =0$ case for which $B$ is a
subspace, and so $\dim A=1$, this notion coincides
with the private quantum channel \cite{AMTW00} and private subspace \cite%
{BHS05,BRS04}. Note that if the definition is satisfied for
$\mathcal{M}_{A^{\prime }|A}$ and $\mathcal{D}_{B^{\prime }|B}$
where the latter is a deletion channel that maps all states on $B$
to a \emph{mixed }state $\omega _{B^{\prime }},$ then we can
always define $\mathcal{M}_{S^{\prime }|A}^{\prime
}=\mathcal{M}_{A^{\prime }|A}\otimes \omega _{B^{\prime }}$ and a
\emph{pure }deletion channel $\mathcal{D}_{\mathbb{C}|B}=\mathrm{
tr}_{B}$ such that the definition is satisfied. Consequently, the
definition of a private subsystem could equally well specify that
$\mathcal{E}\circ \mathcal{P}_{AB}$ be $\epsilon$-close to a
channel of the form $\mathcal{M}'_{S'|A}\otimes \mathrm{tr}_B$.

The use of diamond norms in quantum computing motivates the
following definition for approximately correctable codes. The
$\epsilon = 0$ case was introduced in \cite{KLP05,KLPL05}.
\begin{definition}
    \label{define:correctable}
        Given $\epsilon \geq 0$, we say $B$ is an $\epsilon
        $-\emph{correctable subsystem} for $\mathcal{E}_{S^{\prime }|S}$
        if there is a channel $\mathcal{ R}_{S|S^{\prime }}$ and a channel
        $\mathcal{N}_A$ such that
        \begin{equation}
            \label{eq:correctable}
                \left\Vert \mathcal{R}_{S|S^{\prime }}\circ \mathcal{E}_{S^{\prime
                }|S}\circ \mathcal{P}_{AB}-\mathcal{N}_A\otimes
                \textrm{id}_B\right\Vert _{\Diamond }\leq \epsilon .
        \end{equation}
        where $\textrm{id}_B$ is the identity channel on $B$.
        If Eq.~(\ref{eq:correctable}) holds with $\epsilon=0$, we say that $B$ is a {\em correctable subsystem}.
\end{definition}

Finally, we define the notion of a complementary pair of channels,
which has arisen recently in the analysis of channel capacity
problems \cite {DS03,Hol06,KMNR05} and a continuity theorem for
the Stinespring dilation \cite{KSW06,KSW07}.
\begin{definition}
    \label{define:complement}
        Let $\mathcal{E}_{S^{\prime }|S}$ and $\mathcal{E}%
        _{S^{\prime \prime }|S}^{\sharp}$ be channels on a system $S$ with output spaces
        $S^{\prime }$ and $S^{\prime \prime }$ respectively. Then we say $\mathcal{E}
        $, $\mathcal{E}^{\sharp}$ form a \emph{complementary pair} if there is an
        isometric channel $\mathcal{V}_{S^{\prime }S^{\prime \prime }|S}$ such that
        \begin{equation}
            \label{complementary}
                \mathcal{E}_{S^{\prime }|S}={\mathrm{tr}}_{S^{\prime \prime }}\circ \mathcal{%
                V}_{S^{\prime }S^{\prime \prime }|S} \, , \quad
                \mathcal{E}_{S^{\prime \prime }|S}^{\sharp}={\mathrm{tr}}_{S^{\prime }}\circ \mathcal{V}%
                _{S^{\prime }S^{\prime \prime }|S}.
        \end{equation}
\end{definition}
The Hilbert space $S''$ (respectively $S'$) is a {\it dilation
space} for $\mathcal{E}$ (respectively $\mathcal{E}^{\sharp}$),
and $\mathcal{V}_{S^{\prime }S^{\prime \prime }|S}$ is an {\it
isometric dilation} of both. Complementary pairs arise frequently
in quantum information theory. As a consequence of the Stinespring
Dilation Theorem \cite{Sti55}, every channel may be seen to arise
from an environment Hilbert space $E$ (of dimension at most the
product of the input and output Hilbert space dimensions if the
dilation is minimal), a pure state $|\psi
\rangle $ on the environment, and a unitary operator $U$ on the composite $%
SE $ in the following sense: $\mathcal{E}(\sigma )={\mathrm{tr}}_{E}\big(%
\mathcal{U}(\sigma \otimes |\psi \rangle \!\langle \psi |)\big).$
Tracing
out the system instead yields a complementary channel: $\mathcal{E}%
^{\sharp}(\sigma )={\mathrm{tr}}_{S}\big(\mathcal{U}(\sigma
\otimes |\psi \rangle \!\langle \psi |)\big).$ The corresponding
isometric
form is $\mathcal{E}%
^{\sharp}(\sigma
)={\mathrm{tr}}_{S}\big(\mathcal{V}(\sigma)\big),$ where
$\mathcal{V}$ is implemented by the isometry $V |\phi \rangle = U
|\phi \rangle \, |\psi \rangle$. A simple example is useful in
illustrating the concept.

\begin{lemma}\label{complementlemma}
The identity channel and the trace channel form a complementary
pair.
\end{lemma}

The proof is straightforward. A dilation space for $\mathrm{id}_S$
need only be one-dimensional, $E=\mathbb{C},$ and an isometric
dilation $V$ may be chosen to be simply multiplication by a phase
factor. The complement defined by this dilation is simply
$\mathrm{tr}_S$. We now state our main result.

\begin{theorem}
\label{csps} Let $\mathcal{E}$ and $\mathcal{E}^{\sharp}$ be
complementary
channels. If a subsystem $B$ is $\epsilon$-correctable (respectively $\epsilon$-private) for $%
\mathcal{E},$ then it is $2\sqrt{\epsilon}$-private (respectively $2\sqrt{\epsilon}$-correctable) for $%
\mathcal{E}^{\sharp}.$ The ideal result, obtained by setting
$\epsilon =0$ implies that $B$ is a correctable subsystem for
$\mathcal{E}$ if and only if $B$ is a private subsystem for
$\mathcal{E}^{\sharp}$.
\end{theorem}

The key technical device in the proof is the continuity theorem of
\cite{KSW06}, which we state for completeness.
\begin{theorem}
\label{cont} Let $\mathcal{E}$, $\mathcal{E}^{\prime} \mathpunct :
\mathcal{L}(X) \rightarrow \mathcal{L}(Y)$ be arbitrary quantum
channels, and let $V$ and $V^{\prime}$ be two corresponding
isometric dilations with a common dilation space $Z$. Then
\begin{equation}
    \label{eq:continuity01}
        \left\Vert \mathcal{E}-\mathcal{E}^{\prime}
        \right\Vert _{\Diamond }
        \leq 2\,\,\mathrm{\min_{U}}\,\left\Vert (I_{Y}\otimes
        U)V-V^{\prime}\right\Vert _{\infty } \, ,
\end{equation}
where the minimum is taken over all unitary $U$ on $Z$. Moreover,
if $\dim Z \geq 2 \, \dim X \dim Y$ we also have
\begin{equation}
    \label{eq:continuity02}
        \mathrm{\min_{U}}\left\Vert (I_{Y}\otimes
        U_{})V-V^{\prime}\right\Vert _{\infty }^{2} \leq \left\Vert
        \mathcal{E}-\mathcal{E}^{\prime}
        \right\Vert _{\Diamond } \, .
\end{equation}
\end{theorem}
\strut We note that the continuity theorem has recently been
extended to completely positive maps between arbitrary $C^{\ast
}$-algebras \cite{KSW07}. This should allow for the extension of
the complementarity theorem from finite-dimensional matrix
algebras
to infinite-dimensional ones.\\
\\

\begin{figure}[h]
\includegraphics[scale=.5]{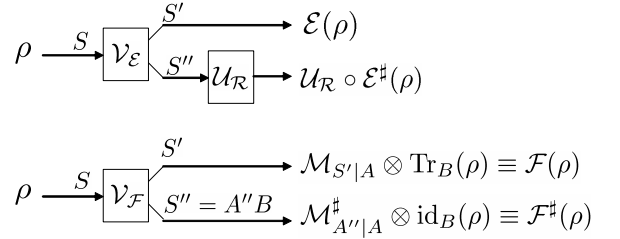}
\caption{Channels involved in the inference from private to
correctable in Theorem~\ref{csps}.} \label{fig1}
\end{figure}

{\bf Proof of Theorem~\ref{csps}:} It suffices to prove the result
in the case $S = A\otimes B$. The general case can then be
obtained by considering the restricted channels $\mathcal E
\circ\mathcal P_{AB}$ and $\mathcal E^\sharp \circ \mathcal
P_{AB}$ (which are complementary). Let $V_{\mathcal{E}}:S
\rightarrow S'\otimes S''$ be the dilation isometry for
$\mathcal{E}$ and $\mathcal{E}^\sharp$. First, suppose $B$ is
$\epsilon$-private for $\mathcal E$ in the sense of
Def.~\ref{define:private}. It is then $\epsilon$-close to a
channel $\mathcal{F} := \mathcal M_{S'|A} \otimes \mathrm{tr}_B.$
By possibly enlarging the dilation spaces of $\E$ and
$\mathcal{F}$, we may always assume without loss of generality
that these spaces are isomorphic to one another and satisfy the
dimension bounds of Theorem~\ref{cont}.
Eq.~(\ref{eq:continuity02}) then guarantees the existence of a
unitary $U_\mathcal{R}$ on the common dilation space $S''$ such
that
\begin{equation}
\label{leftsideeqn} || ( U_\mathcal{R} \otimes I_{S'}) \,
V_{\mathcal{E}} - V_\mathcal{F} ||_\infty
\, \leq \, ||\mathcal{E} - \mathcal{F} ||_{\diamond%
}^{1/2} \leq \sqrt{\epsilon} \, .
\end{equation}
Define $\F^{\sharp}:= \mathrm{tr}_{S'}\circ V_{\mathcal{F}}$. By
Eq.~(\ref{eq:continuity01}), we infer that
\begin{equation}
\label{rightsideeqn}
    || \mathcal{U}_{\mathcal{R}}\circ\mathcal{E}^\sharp -
    \mathcal{F}^\sharp ||_{\diamond} \leq
    2|| ( U_\mathcal{R} \otimes I_{S'}) \, V_{\mathcal{E}} -
V_\mathcal{F} ||_\infty .
\end{equation}
Define $A''$ by $S''=A''\otimes B$.  Note that by
Lemma~\ref{complementlemma},
$\mathcal{F}^{\sharp}=\M^{\sharp}_{A''|A} \otimes \mathrm{id}_B$
where $\M^{\sharp}_{A''|A} := \mathrm{tr}_{S'}\circ
V_{\mathcal{M}}$ and $V_{\mathcal{M}}$ is an isometric dilation of
$\mathcal{M}_{S'|A}$ with dilation space $A''$.  Finally, define
$A_0$ by $A''=A_0\otimes A$, and define the channels
$\mathcal{N}_A := \mathrm{tr}_{A_0}\circ \M^{\sharp}_{A''|A}$ and
$\mathcal{R} := \mathrm{tr}_{A_0}\circ \mathcal{U}_{\mathcal{R}}$.
Tracing over $A_0$ in the left-hand side of
Eq.~(\ref{rightsideeqn}), noting that the diamond norm is
nonincreasing under partial trace, and using
Eq.~(\ref{leftsideeqn}), we obtain
\begin{equation}
    \label{eq:bound}
        \cb{\R \circ \E^{\sharp} - \mathcal{N}_A  \otimes \id_{B}} \leq 2 \sqrt{\epsilon} \, ,
\end{equation}
which implies that $B$ is $2 \sqrt{\epsilon}$-correctable for
$\E^{\sharp}$. $\blacktriangle$

Suppose now that $B$ is $\epsilon$-correctable for $\mathcal E$ in
the sense of Def.~\ref{define:correctable}, so that there exists a
channel $\mathcal{R}$ such that $\mathcal{R}\circ\mathcal E$ is
$\epsilon$-close to a channel $\G := \mathcal{N}_A \otimes
\mathrm{id}_B$.  Again, we may assume that the dilation spaces of
$\mathcal{R}\circ\mathcal E$ and $\G$ are isomorphic and satisfy
the dimension bounds of Theorem~\ref{cont}.  If we denote the
dilation spaces of $\E$ and $\mathcal{R}$ by $S''$ and $S_0$
respectively, then the common dilation space of
$\mathcal{R}\circ\mathcal E$ and $\G$ is $S_0 \otimes S''$.
Letting $V_{\mathcal{E}}$, $V_{\mathcal{R}}$ and $V_{\mathcal{G}}$
denote the isometric dilations of $\E$, $\mathcal{R}$  and
$\mathcal{G}$, we infer from Eq.~(\ref{eq:continuity02}) that
there exists a unitary $U$ on $S_0 \otimes S''$ such that
\begin{equation}
    \label{leftsideeqn1}
        || ( V_\mathcal{R} \otimes I_{S''}) \, V_{\mathcal{E}} - (I_{S}\otimes U) V_
        \mathcal{G} ||_\infty \leq ||\mathcal{R}\circ \mathcal{E} - \mathcal{G} ||^{1/2} \leq \sqrt{\epsilon} \, .
\end{equation}
Define $\G^{\sharp}:= \mathrm{tr}_{S}\circ
\mathcal{V}_{\mathcal{G}}$ and $\mathcal{R}^{\sharp} :=
\mathrm{tr}_{S}\circ \mathcal{V}_{\mathcal{R}}$. By
Eq.~(\ref{eq:continuity01}), we infer that
\begin{equation}
\label{rightsideeqn1}
    || (\mathcal{R}^{\sharp}\otimes \mathrm{id}_{S''})\circ
    \mathcal{V}_{\mathcal{E}}
    - \mathcal{U} \circ \mathcal{G}^{\sharp} ||_{\diamond} \leq
    2\sqrt{\epsilon} \, .
\end{equation}
By Lemma~\ref{complementlemma}, we have
$\mathcal{G}^{\sharp}=\mathcal{N}^{\sharp}_{S_0 S''|A} \otimes
\mathrm{tr}_B$ where $\mathcal{N}^{\sharp}_{S_0 S''|A} :=
\mathrm{tr}_{S}\circ \mathcal{V}_{\mathcal{N}}$ and where
$V_{\mathcal{N}}$ is an isometric dilation of $\mathcal{N}_A$.
Finally, if we define $\mathcal{M}_{S''|A}:= \mathrm{tr}_{S_0}
\circ\mathcal{N}^{\sharp}_{S_0 S''|A}$, and take the trace over
$S_0$ on the left-hand side of Eq.~(\ref{rightsideeqn1}) (noting
that the diamond norm is nonincreasing under partial trace), we
find
\begin{equation}
    \label{eq:comp02}
        \cb{\E^{\sharp} - \mathcal{M}_{S''|A} \otimes \mathrm{tr}_B} \leq 2 \, \sqrt{\epsilon} \, .
\end{equation}
Hence, $B$ is $2 \sqrt{\epsilon}$-private for $\E^\sharp$, as
claimed. $\rule{0.5em}{0.5em}$

\begin{figure}[h]
\includegraphics[scale=.5]{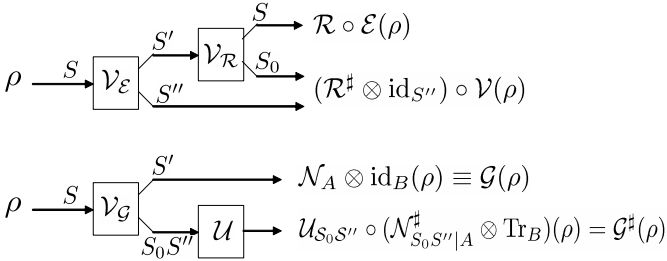}
\caption{Channels involved in the inference from correctable to
private in Theorem~\ref{csps}.} \label{fig2}
\end{figure}

As a simple example for the ideal case, consider a two-qubit noise
model that induces a phase flip $Z_1$ on the first qubit with
probability one half. The associated channel on $\mathbb C^2
\otimes \mathbb C^2$ is $\E(\sigma) = \frac{1}{2}( \sigma + Z_1
\sigma Z_1)$. The code subspace $\C$ with basis
$\{\ket{00},\ket{01}\}$ is a decoherence-free subspace for $\E$,
in the sense that $\E(\sigma) = \sigma$ for all $\sigma$ supported
on the code space $\C$. The map $\E$ can be obtained by tracing
out a single qubit environment $E$ as $ \E(\sigma) = \Tr_E ( \, U
(\sigma \otimes \kb{0}{0})U^\dagger \,), $ where $U$ is the
unitary $ U \propto  \one_2\otimes\kb{0}{0} + Z_1 \otimes\kb{0}{1}
+ \one_2\otimes\kb{1}{0}  -Z_1 \otimes\kb{1}{1}. $ Direct
computation reveals the complementary channel $ \E^\sharp(\sigma)
= \Tr_S ( \, U (\sigma \otimes \kb{0}{0})U^\dagger \, ),$
satisfies $\E^\sharp(\sigma) = \Tr(\sigma) \rho_1 + \Tr(\sigma
Z_1) \rho_2$, where $\rho_1 \propto \kb{0}{0}+\kb{1}{1}$ and
$\rho_2 \propto \kb{0}{1}+\kb{1}{0}$. Theorem~\ref{csps} predicts
the messenger space $\C$ is a private subspace for $\E^\sharp$.
Indeed, one can easily verify that for all $\sigma$ supported on
the code space $\C$ we have $\E^\sharp(\sigma) = \Tr(\sigma)\, P$,
with the projector
$P = \rho_1 + \rho_2$.\\

Quantum secret sharing provides a nice example of the utility of
the complementarity theorem. A $((k,n))$ threshold scheme for
quantum secret sharing is a protocol that encodes the quantum
state of a system $S$ (the quantum secret) into $n$ systems, one
held by each party, such that $k$ parties or more can recover the
secret, while $k-1$ or fewer cannot gain any information about it
\cite{CGL99}. Our result demonstrates that one can achieve a
scheme that approximates the ideal functionality as follows: the
reduction of the encoding map to any $k$ or more parties is
$\epsilon$-correctable while to any $k-1$ or fewer parties it is
$2\sqrt{\epsilon}$-private. As long as the encoding map is an
isometry, then by our theorem and the definition of complementary
maps, if the input space is $\epsilon$-correctible for the
reduction of the encoding map to any $k$ or more parties, then it
is $2\sqrt{\epsilon }$-private for the reduction to any $n-k$ or
fewer parties.  Therefore, as long as $k-1=n-k,$ or $n=2k-1,$ we
obtain the desired approximation to ideal functionality. This is
the generalization of Corollary 9 of \cite{CGL99}. Furthermore,
every nonisometric encoding among $n$ parties can be understood as
some isometric encoding among $n'>n$ parties where the extra
$n'-n$ shares are discarded. Given that $n^{\prime }=2k-1,$ we
infer that $n<2k-1$. Therefore, the approximation to ideal
functionality described above is impossible if $n\geq 2k$. This is
the generalization of Theorem 2 of \cite{CGL99}.

Our result also finds a simple application in the standard
paradigm of quantum communication where it is presumed that any
dilation space for the channel $\mathcal{E}$ linking Alice to Bob
ends up in the hands of an adversary. The theorem then implies
that any subsystem that is $\epsilon$-correctable for Bob is
$2\sqrt{\epsilon}$-private for the adversary.

\textit{Acknowledgements.}  We are grateful to the Banff
International Research Station for kind hospitality. D.K. is
grateful for generous support from Deutscher Akademischer
Austauschdienst (DAAD). D.W.K. acknowledges support from NSERC,
ERA, CFI, and OIT. R.W.S. acknowledges support from the Royal
Society.


\end{document}